\documentclass[aps,prl,a4paper,twocolumn,showpacs,showkeys,preprintnumbers,amsmath,amssymb]{revtex4}
\usepackage{graphicx}
\usepackage{bm}
\usepackage{dcolumn}
\usepackage{color}
\usepackage{pst-plot}
\renewcommand{\texttt}{{}}
\newcommand{\be}{\begin{eqnarray}}
\newcommand{\ee}{\end{eqnarray}}
\usepackage{graphicx}

\begin{document}

\title{Black holes in an ultraviolet complete quantum gravity} 
\author{Leonardo Modesto}
\thanks{Electronic address: lmodesto@perimeterinstitute.ca}
\affiliation{Perimeter Institute for Theoretical Physics, 31 Caroline St., Waterloo, ON N2L 2Y5, Canada}

\author{John W. Moffat}
\thanks{Electronic address: 	john.moffat@utoronto.ca}
\affiliation{Perimeter Institute for Theoretical Physics,
31 Caroline Street North, Waterloo, ON N2L 2Y5, Canada\\
Department of Physics and Astronomy,
University of Waterloo, Waterloo, ON N2L 3G1 Canada}

\author{Piero Nicolini}
\thanks{Electronic address: nicolini@th.physik.uni-frankfurt.de}
\affiliation{Frankfurt Institute for Advanced Studies (FIAS) and 
Institut f\"ur Theoretische
Physik, Johann Wolfgang Goethe-Universit\"at, Ruth-Moufang-Strasse 1, 
 D-60438
Frankfurt am Main, Germany}

\date{\small\today}

\begin{abstract} \noindent
In this Letter we derive the gravity field equations by varying the action for an ultraviolet complete quantum gravity. Then we consider the case of a static source term and we determine an exact black hole solution. As a result we find a regular spacetime geometry: in place of the conventional curvature singularity extreme energy fluctuations of the gravitational field at small length scales provide an effective cosmological constant in a region locally described in terms of a deSitter space. We show that the new metric coincides with the noncommutative geometry inspired Schwarzschild  black hole.  Indeed we show that the ultraviolet complete quantum gravity, generated by ordinary matter is  the dual theory of ordinary Einstein gravity coupled to a noncommutative smeared matter. In other words we obtain further insights  about that quantum gravity mechanism which improves Einstein gravity in the vicinity of curvature singularities. This corroborates all the existing literature in the physics and phenomenology of noncommutative black holes.
\end{abstract}

\maketitle
\noindent
An ultraviolet (UV) complete quantum gravity theory has been formulated using a diffeomorphism invariant action
in which the gravitational strength is
\begin{equation}
\sqrt{G (x)}=\sqrt{G_N}{\cal F}\left(\Box(x)/\Lambda_G^2\right),
\label{gcoord}
\end{equation}
where $G_N$ is Newton's constant,
$\Box=g^{\mu\nu}\nabla_\mu\nabla_\nu$ is the generally covariant D'Alembertian operator,
and ${\cal F}$ is an entire function~\cite{Moffat:2010bh}.
Moreover, $\Lambda _G$ is a constant gravitational energy scale and the entire function ${\cal F}$ has no poles in the finite complex plane. The quantum gravity perturbation theory expanded against a
fixed Minkowski background spacetime is locally gauge invariant and unitary to all
orders. The graviton-graviton and graviton-matter loops in Euclidean momentum space
are finite to all orders. The graviton tree graphs are point-like and local
maintaining the macroscopic local and causal property of gravity.

The attempts to use noncommutative geometry to deduce phenomenological results from
a perturbative expansion in the noncommutative parameter $\theta$, run into the
difficulty that a truncation of the Moyal $\star$-product makes the theory local and leads
to a lack of renormalizabilty in the quantum gravity version of the theory \cite{Chamseddine:1992yx} (for a general review on the topic see \cite{ncgreview}).  In
contrast, we find that the nonlocal nature of the vertex function in the
perturbative UV complete quantum gravity theory does not require a truncation,
retaining its full nonlocality while the graviton is described by a local,
microcausal field and propagator.  The same holds true for the UV complete standard
model Feynman rules in which the interaction of particles is nonlocal but the
physical fields and propagators are local and causal~\cite{Moffat2}.


In the following, we investigate the consequences of the UV complete
quantum gravity for black holes. Along the lines in \cite{Moffat:2010bh} we start
with the four-dimensional action for gravity:
\begin{eqnarray}
S_{\mathrm{grav}} 
=\frac{1}{ 16\pi} \int \, d^{4} x \, \sqrt{-g} \, G^{-1}(x)
\left(R- 2\lambda \right),
\label{uvfeh}
\end{eqnarray}
where we use the signature $(-+++)$ and $\lambda$ is the cosmological constant.
The action (\ref{uvfeh}) has a nonlocal character because of the presence of the term ${\cal F}^{-2}$.
In the Euclidean momentum space representation:
$\sqrt{G (p^2)}=\sqrt{G_N}{\cal F}\left(p^2/\Lambda_G^2\right)$
and in addition we require the on shell condition $G(0)=G_N$.
The field equations are obtained by varying the action (\ref{uvfeh})
with respect to the metric $g_{\mu\nu}$. By neglecting surface terms
coming from the variation of the generally-covariant D'Alembertian \cite{Mureika:2010je}, we find
\begin{eqnarray}
{\cal F}^{-2}\left(\Box(x)/\Lambda_G^2\right)\left(R_{\mu \nu} - \frac{1}{2} g_{\mu \nu}R \right)= 8 \pi
G_N T_{\mu \nu},
\label{e0}
\end{eqnarray}
where  we have set $\lambda=0$. We notice that (\ref{e0}) can be cast in a different form by ``shifting'' the the operator ${\cal F}^{-2}$ to the
 r.h.s. leaving the l.h.s. in the canonical form, i.e.,
\begin{eqnarray}
R_{\mu \nu} - \frac{1}{2} g_{\mu \nu}R = 8 \pi
G_N{\cal S}_{\mu\nu},
\label{e1}
\end{eqnarray}
where the tensor
\begin{equation}
{\cal S}_{\mu\nu}\equiv{\cal F}^2\left(\Box(x)/\Lambda_G^2\right) T_{\mu \nu}.
\end{equation}
We notice that the new source term is conserved, i.e., $\nabla_\mu S^{\mu\nu}=0$.
As a matter of fact, (\ref{e1}) describes Einstein gravity coupled to a
generalized matter source term, while (\ref{e0}) describes the UV complete quantum gravity produced by ordinary
matter. The two interpretations are physically equivalent.

\begin{figure}[ht]
\begin{center}
\includegraphics[width=4.5cm,angle=0]{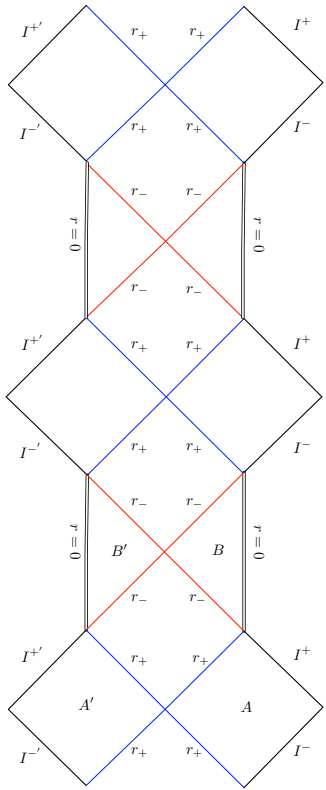}
\caption{\label{penrose}The solution admits one, two or no horizons depending on $M$. In the case of two horizons,  $f(r_\pm)=0$, the Penrose diagram resembles the Reissner-Norstr\"{o}m geometry, except for the origin where a regular deSitter core lies in place of the curvature singularity.}
\end{center}
\end{figure}

Our main purpose is to solve the field equations by assuming
a static source, i.e., the four-velocity field $u^\mu$ has only
a non-vanishing time-like component
$u^\mu\equiv ( u^0 , \vec{0} )$ 
$u^0= (- g^{00})^{-1/2}$.
The component  $T^0\,_0$ of the energy-momentum tensor for a static source is given by \cite{DeBenedictis:2007bm}
\begin{eqnarray}
 T^0\,_0=-\frac{M}{4\pi\, r^2}\delta ( r ). \label{t00}
\end{eqnarray}
The metric of our spacetime is assumed to be given by the usual static, spherically symmetric form
\begin{equation}
ds^2=- f(r)dt^2 + \frac{dr^2}{f(r)}+r^2\Omega^2,
\end{equation}
where
\begin{equation}
f(r)=1-\frac{2G(r) M }{r}.
\end{equation}
In Einstein gravity $G(r)=G_N$ and one obtains the Schwarzschild geometry. To solve field equations we follow the form (\ref{e1}), by determining the generalized matter source term ${\cal S}_{\mu\nu}$.
The metric component can be written as
\begin{equation}
f(r)=1-\frac{2G_N \,  m(r)}{r} ,
\label{f}
\end{equation}
where
\begin{equation}
m(r) = - 4\pi \int dr r^2\ {\cal S}^0\,_0 .
\label{mass}
\end{equation}
For later convenience we temporarily adopt free falling Cartesian-like coordinates and we calculate
\begin{equation}
{\cal S}^0\,_0= - M {\cal F}^2\left(\Box(x)/\Lambda_G^2\right)\delta(\vec{x})\equiv - \rho_{\Lambda_G}(\vec{x}).
\end{equation}
The covariant conservation and the additional condition, $g_{00}=-g_{rr}^{-1}$,
completely specify the form of ${\cal S}^\mu{}_\nu$.  Before proceeding further, we need to specify the form of  ${\cal F}$ within the class of entire functions.  We do not know the unique choice. However, a simple form of ${\cal F}$ fulfilling the properties we require is
\begin{equation}
{\cal F}\left(p^2\right)=\exp\left(\frac{-p^2}{2\Lambda_G^2}\right)
\label{natural}
\end{equation}
in Euclidean momentum space \cite{Moffat:2010bh}. As a check of consistency we can see that all Feynman graviton loops containing at least one vertex function ${\cal F}$  are ultraviolet finite. As a consequence we have
\begin{eqnarray}
&&
{\cal F}^2\left(\Box(x)/\Lambda_G^2\right)\delta(\vec{x}) = \nonumber \\
&& = e^{\nabla^2/\Lambda_G^2}\delta(\vec{x}) =
\frac{1}{(2\pi)^3} \int d^3p \ e^{-p^2/\Lambda_G^2} e^{i\vec{x}\cdot\vec{p}}. 
\end{eqnarray}
By calculating the above integral, one gets
\begin{equation}
\rho_{\Lambda_G}(\vec{x})=M\left(\frac{1}{2}\sqrt{\frac{\Lambda_G^2}{\pi}}\right)^3 e^{-\vec{x}^2\Lambda_G^2/4}.
\end{equation}
We notice that the generalized matter energy density profile is a Gaussian whose width is $1/\Lambda_G$. This means that for energies smaller than $\Lambda_G$ the function $\rho_{\Lambda_G}(\vec{x})$ approaches the Dirac delta distribution $\delta(\vec{x})$. This is equivalent to say that the function $m(r)$ becomes the total mass $M$ in Newtonian gravity, since we are probing the system at asymptotic length scales where the UV complete quantum gravity is nothing but Einstein gravity.
The final step is to obtain the mass function of the matter.
From (\ref{mass}) one finds
\begin{equation}
m(r)=M\, \left[ 1- \frac{\Gamma\left(3/2; r^2\Lambda_G^2/4\right)}{\Gamma(3/2)} \right],
\label{mass2}
\end{equation}
where
\begin{equation}
\Gamma\left(3/2; r^2\Lambda_G^2/4\right)=\int_{r^2\Lambda_G^2/4}^\infty\ dt \, t^{1/2}\ e^{-t}
\end{equation}
and $\Gamma(3/2)=\sqrt{\pi}/2$ is Euler's gamma function.
By expanding (\ref{mass2}) for $r\gg 1/\Lambda_G$ we have 
\begin{equation}
m(r)\approx M \ \left[ 1- \frac{\Lambda_G}{\sqrt{\pi}}\ r \, e^{-r^2\Lambda_G^2/4}\right],
\end{equation}
which matches the required value $M$ up to exponentially suppressed corrections. Such corrections are important since the UV complete quantum gravity can lead to experimentally testable deviations from Newton's law \cite{Nicolini:2010nb}
\begin{equation}
\phi_N(r)=G_{N}\frac{M}{r}\left[ 1- \frac{\Lambda_G}{\sqrt{\pi}}\ r\ e^{-r^2\Lambda_G^2/4}\right].
\end{equation}
 On the other hand we can observe the UV completeness of the theory at work in the high energy regime. Indeed if we expand (\ref{mass2}) for $r\ll 1/\Lambda_G$ we get
\begin{equation}
m(r)\approx \frac{1}{6}\frac{M}{\sqrt{\pi}}\ r^3 \ \Lambda_G^3.
\end{equation}
At this point we can substitute this value into (\ref{f}) to get
\begin{eqnarray}
\hspace{-0.15cm} ds^2\approx - \left(1-\frac{1}{3}\Lambda_{\mathrm{eff}} r^2\right)dt^2 + \frac{dr^2}{\left(1-\frac{1}{3}\Lambda_{\mathrm{eff}} r^2\right)}+r^2\Omega^2.
\end{eqnarray}
This is a deSitter line element whose effective cosmological constant
$\Lambda_{\mathrm{eff}} = M   G_N  \Lambda_G^3/\sqrt{\pi}$,
accounts for the ``vacuum energy'' of the ``field'' $g_{\mu\nu}$. In other words we show that in the UV complete quantum gravity the gravitational field acquires a repulsive character as far as one probes the seething fabric of spacetime. Again we can say that the intrinsic nonlocality of the action (\ref{uvfeh}) is able to tame the curvature singularity of the Schwarzschild solution (see the Fig. \ref{penrose}). By calculating curvature tensors at the origin one finds that they are finite. For instance the Ricci scalar reads
\begin{equation}
R(0)=\frac{4M G_N  \Lambda_G^{3}}{\sqrt{\pi}}.
\end{equation}
\begin{figure}[ht]
\begin{center}
\includegraphics[width=6.0cm,angle=0]{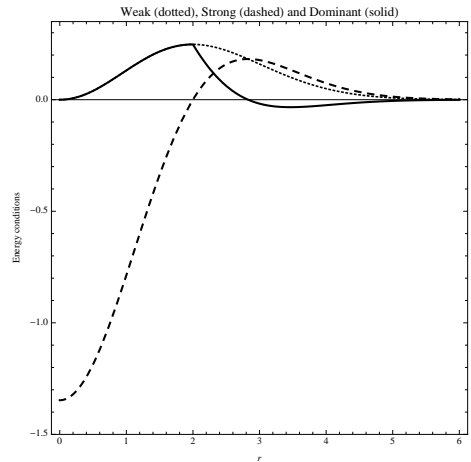}
\caption{\label{SEC} The dashed curve is the function $(\rho_{\Lambda_G}
+p_r+2p_\perp) / \Lambda^4_G$ vs $r\Lambda_G$ (strong energy condition); the solid curve is $(\rho_{\Lambda_G}-|p_\perp|) / \Lambda^4_G$ (dominant energy condition); the dotted curve is $(\rho_{\Lambda_G}+p_\perp) / \Lambda^4_G$ (weak energy condition). In a region within $r=6/\Lambda_G$ all conditions are violated
.
}
\end{center}
\end{figure}

To get more insights about the nature of the generalized matter that generates the above regular geometry, it is worthwhile to analyse the energy conditions. First one has to determine the nonvanishing pressure terms coming from the conservation of the stress tensor ${\cal S}_{\mu\nu}$. For instance the strong energy condition reads
\begin{equation}
\rho_{\Lambda_G} +p_r+2p_\perp \sim e^{-r^2\Lambda^2_G/4}\left(\frac{r^2\Lambda^2_G}{2}-2\right)\ge 0,
\end{equation}
where $p_r$ is the radial pressure and $p_\perp$ is the angular one. From Fig. \ref{SEC} we can see that in the vicinity of the origin the matter has an exotic character, i.e., strong, dominant and weak energy conditions are violated.

At this point we could proceed further by studying the horizon equation, the thermodynamic properties and the global structure of the solution. However we prefer to stop here, since the line element we have found
\begin{equation}
f(r)=1-\frac{2G_N M}{r}\ \frac{\gamma\left(3/2; r^2\Lambda_G^2/4\right)}{\Gamma(3/2)}
\label{f2}
\end{equation}
is nothing but the noncommutative geometry inspired Schwarzschild black hole \cite{Nicolini:2005vd}, where $\gamma\left(3/2;\, x\right)=\Gamma\left(3/2\right)-\Gamma\left(3/2; \, x\right)$. In this scenario the noncommutative geometry induced minimal length $\sqrt{\theta}$ is nothing but $1/\Lambda_G$. The above metric was derived by one of us and his coworkers Smailagic, Spallucci after a long path. At the time there were already several attempts of incorporating noncommutative effects in black hole physics. All such attempts were based on expansions of the $\star$-product among vielbein fields entering gravity Lagrangians
\cite{Chamseddine:1992yx}. The problem is that any truncation at a desired order in the noncommutative parameter basically destroys the non-locality encoded in the $\star$-product and gives rise to a local theory, plagued by spurious momentum-dependent terms. As a result, in spite of the mathematical exactitude, all the proposed corrections coming from this kind of approach failed in curing the bad short distance behavior of black hole solutions in Einstein gravity \cite{Chaichian:2007we}.
Against this background, the noncommutative geometry inspired Schwarzschild solution was derived in an \textit{effective} way. 

Instead of embarking on the interesting but difficult problem of formulating a computationally viable noncommutative gravity, it is worthwhile to study the average effect of manifold noncommutative fluctuations on point like sources. In a series of papers based on the use of coordinate coherent states \cite{Smailagic:2003yb,Smailagic:2003rp,Smailagic:2004yy,Spallucci:2006zj} (and recently confirmed by means of another approach based on Voros products \cite{Banerjee:2009xx}), it has been shown that the mean position of a pointlike object in noncommutative geometry is no longer governed by a Dirac delta function, but by a Gaussian distribution.

As a second step toward the solution  (\ref{f2}), it has been shown that primary corrections to any field equation in the presence of a noncommutative smearing can be obtained by replacing the source term (matter sector) with a Gaussian distribution, while keeping formally unchanged differential operators (geometry sector) \cite{Nicolini:2005zi}. In the specific case of the gravity field equations this is equivalent to saying that the only modification occurs at the level of the energy-momentum tensor,  while $G_{\mu\nu}$ is formally left
unchanged.
In this spirit further solutions have been derived corresponding to the case of dirty \cite{dirty}, charged \cite{charged}, spinning \cite{spinning} black holes (for a review see \cite{Nicolini:2008aj}).

Another important feature concerns the new thermodynamics of these black holes. Indeed, even for the neutral solution, the Hawking temperature  reaches a maximum before running a positive heat capacity, cooling down phase towards a zero temperature remnant configuration \cite{ncthermo}. As a consequence, according to this scenario quantum back reaction is strongly suppressed in contrast to conventional limits of validity of the semiclassical approximation in the terminal phase of the evaporation. Furthermore the higher-dimensional solutions \cite{higher,chargedhigher}, due to their attractive properties have been recently taken into account in Monte Carlo simulations as reliable candidate models to describe the conjectured production of microscopic black holes in particle accelerators \cite{Gingrich:2010ed}.

Let us consider further our solution and the relation between UV complete quantum gravity and noncommutative geometry. First, the form of equations (\ref{e0}) tells us that we are working in the framework of UV complete quantum gravity \cite{Moffat:2010bh}. Indeed the matter sector is unchanged, while nonlocal modifications enter the geometry. The dual theory is governed by equation (\ref{e1}), that is based on a generalized matter
energy-momentum tensor, keeping the Einstein tensor in the canonical form. It is now clear that the exotic nature of matter is nothing but a ``seething'' noncommutative character of the source term. More specifically, the duality between the two descriptions holds also at the level of the specific choice of the operator ${\cal F}$. Indeed the natural choice (\ref{natural}), i.e., the simplest form within the class of entire functions, corresponds to the case of primary noncommutative geometry corrections to the manifold, as often advocated in \cite{Smailagic:2003yb,Smailagic:2003rp,Smailagic:2004yy,Spallucci:2006zj}. The virtue of these primary corrections is that they are not the result of a truncation in a perturbative expansion, but are intrinsically nonlocal. The natural duality link between the UV complete quantum gravity and the Einstein field equations with a generalized energy-momentum tensor 
sheds light on the interpretation of this key point.
\begin{acknowledgments}
\noindent P.N. is supported by the Helmholtz International Center for FAIR within the
framework of the LOEWE program (Landesoffensive zur Entwicklung Wissenschaftlich-\"{O}konomischer Exzellenz) launched by the State of Hesse. P.N. would like to thank the Perimeter Institute for Theoretical Physics, Waterloo, ON, Canada for the kind hospitality during the period of work on this project. Research at Perimeter Institute is
supported by the Government of Canada through Industry Canada and by the Province of Ontario through the
Ministry of Research \& Innovation. The authors thank E. Spallucci for valuable comments about the manuscript.
\end{acknowledgments}

\end{document}